# Consistency between household and county measures of K-12 onsite schooling during the COVID-19 pandemic

**Authors:**

Carly Lupton-Smith[1], Elena Badillo Goicoechea[2], Megan Collins[3,4], Justin Lessler[5], M. Kate Grabowski[5], and Elizabeth A. Stuart[1,2]

[1] Department of Biostatistics, Johns Hopkins Bloomberg School of Public Health, Baltimore, MD, USA

[2] Department of Mental Health, Johns Hopkins Bloomberg School of Public Health, Baltimore, MD, USA

[3] Wilmer Eye Institute, Johns Hopkins School of Medicine, Baltimore, MD, USA

[4] Berman Institute of Bioethics, Johns Hopkins University, Baltimore, MD, USA

[5] Department of Epidemiology, Johns Hopkins Bloomberg School of Public Health, Baltimore, MD, USA

**Statement of Research Support:** Funding for this research was supported by the Johns Hopkins University Discovery Award and the Johns Hopkins University COVID-19 Modeling and Policy Hub Award

**Abstract**

The academic, socioemotional, and health impacts of school policies throughout the COVID-19 pandemic have been a source of many important questions that require accurate information about the extent of onsite schooling that has been occurring throughout the pandemic. This paper investigates school operational status data sources during the COVID-19 pandemic, comparing self-report data collected nationally on the household level through a Facebook-based survey with data collected at district and county levels throughout the country. The percentage of households reporting in-person instruction within each county is compared to the district and county data at the state and county levels. The results show high levels of consistency between the sources at the state level and for large counties. The consistency levels across sources support the usage of the Facebook-based COVID-19 Symptom Survey as a source to answer questions about the educational experiences, factors, and impacts related to K-12 education across the nation during the pandemic.

**Introduction**

Going into the fall of 2020 in the midst of the COVID-19 pandemic, schools across the United States had to make a decision about how to educate children and adolescents effectively and safely. Schools varied in their instruction policies from fully online, to hybrid online and onsite, to all onsite, and these policies have often changed multiple times during the course of the pandemic (Honein et al., 2021). Furthermore, these decisions were made at various levels: county, district, school, or even at the household level, as schools often gave parents the choice about whether to send their children in for onsite instruction. The potential implications of these instructional models on children, their families, and their communities are substantial. However, measuring the extent of onsite instruction is complex; how it is done, and the quality of the underlying data, can have significant effects on the conclusions drawn from studies that examine links between school operations and relevant outcomes during and beyond the COVID-19 pandemic.

There are reasons to be very interested and concerned regarding links between school instructional models during the pandemic and COVID-19, social, mental health, and academic outcomes. On the one hand, when schools were closed in March of 2020, evidence indicated that these closures reduced the spread of COVID-19 (Auger et al., 2020) and there have been multiple papers since aiming to study the links between onsite schooling and COVID transmission (e.g., Harris et al., 2021, Goldhaber et al., 2020, Zimmerman et al., 2020, Falk et al., 2021, Honein et al., 2021), with varying definitions of school "reopening" (i.e. all onsite, hybrid) and whether they also consider mitigation measures such as mask-wearing and physical distancing.  On the other hand, beyond COVID infection implications of school instructional modalities, there are also potential large impacts on children and their families in other ways. Food insecurity can increase for families who count on school as a supplier of meals for their children (Van Lancker and Parolin, 2020), and schools are often the sites of routine health care for children, such as vision and hearing screenings and even vaccinations. Furthermore, students with low socioeconomic status might experience difficult conditions for online learning. Previous research has also shown, for this group of students, that summer vacations lead to a loss of academic achievement compared to students with higher socioeconomic status (Van Lancker and Parolin, 2020). There are reasons to expect a similar dynamic during pandemic-related school closures (Eyles et al., 2020). Thus, these school closures have potential to exacerbate disparities across areas and households in the country (Armitage and Nellums, 2020).

There are therefore many factors that school leaders have had to consider when making school policy decisions during this pandemic, including the instruction modality and also any preventive measures put in place for onsite instruction. And until Centers for Disease Control and Prevention guidance in February 2021, there was little formal guidance from the federal government on school reopening, leading to substantial variability in these decisions across the country (CDC, 2021). Little work has been done so far to understand these processes and decisions. One study found that mass partisanship and the strength of the teacher's union were the strongest predictors of a district's decision to reopen or not reopen schools (Hartney and Finger, 2020). Interestingly, the severity of COVID-19 in the area was a potential predictor in their model but was not identified as having a strong relationship.

A challenge in studying the links between school operational status and key outcomes of interest is the variability in operations across the country and the lack of standardized data on the topic. A few sources exist (very few on truly national data), and they vary in how they collect data, how many areas in the country they report on, how often they update these reports, how they define operational status, and to which level of measurement (i.e., district, county) they aggregate. In addition, a limitation of most sources is that they are measured only at the district or county level and do not indicate individual household decisions about onsite instruction. Since many districts offer families a choice to stay fully virtual during the COVID-19 pandemic even if there are onsite options, it is crucial to obtain individual household information about schooling behaviors in order to fully understand the extent of onsite instruction and how it varies across districts (Harris et al, 2021). It is also crucial to understand how well that household self-report relates to district-level policies. In this paper, we thus compare district and county-level instructional status information with a large-scale national daily survey with household-level responses about schooling behaviors. These comparisons help validate the use of a large novel dataset describing COVID-19 educational experiences at the household level. Using this data can provide better information about the decision making of households and the factors involved in those decisions, as well as the effects of different schooling policies.

**Data and Methods**

This section details the household schooling behavior data available from the CMU/Facebook COVID-19 Symptom Survey as well as two other data sources that provide aggregate data at the district or county level, which are used for comparison. Table 1 provides details of the data sources, including the specific questions used to assess instructional modalities.

*CMU/Facebook COVID-19 Symptom Survey*

The CMU/Facebook COVID-19 Symptom Survey is a daily cross-sectional survey that invites a new stratified random sample of adult Facebook users to take the survey each day (Badillo-Goicoechea et al, 2020; Barkay et al, 2020). The survey has a United States version run by Carnegie Mellon University (CMU) and a global version run by the University of Maryland, both in partnership with Facebook (Delphi Group, 2020; University of Maryland, 2020). The current study uses the United States data from CMU/Facebook, which began collection on April 6, 2020 and has had approximately 50,000 respondents every day since. The survey includes questions about COVID-19 symptoms and testing, health-related behaviors such as mask wearing, and mental health; it also gets an individual's zip code to obtain US county (Badillo-Goicoechea et al, 2021, Kreuter et al, 2020). Weights are used to reflect distributions at the state level, specifically a two stage process that involves first weighting by an inverse propensity score to account for non-response bias, and second post-stratification weighting to adjust the Facebook user age and gender distribution to that of the general population (Barkay et al., 2020; Badillo-Goicoechea et al., 2021).

Questions regarding K-12 students' educational experiences were added in Wave 5 of the survey, which began on November 24, 2020 (Delphi Group, 2020). Data used in this study were thus from November 28th, 2020 to January 29th, 2021, with data between December 22nd, 2020 to January 2nd, 2021 removed under the assumption that most schools were on winter break. Data from Puerto Rico and anywhere outside of the US were removed. Analyses of the CMU/Facebook data were restricted to participants who responded that they had children in their household in grades Pre-kindergarten through 12th grade. The full sample had 510,424 respondents.

The symptoms survey asked participants who had school-aged children if any of the following applied to children in the household (Pre-K—grade 12): going to onsite classes full-time, going to onsite classes part-time (Delphi Group, 2020). These responses were grouped into a category of "any onsite school", which was the case if an individual said any children were either going to onsite classes full-time or part-time, or no onsite school, which was the case if both responses were marked "No". For comparison with other data sources, these measures were summarised at the state level using the weights provided by the survey (Barkay et al., 2020). For comparison at the county level, the same survey-provided weights were used, and only survey participants from the 460 most populous counties in the country were included. This allowed for more accurate comparison with other data sources, as smaller counties had very small sample sizes from the CMU/Facebook survey. Therefore, the data for county-level comparisons consisted of 354,548 total respondents. This information is summarized in Table 1.

*Burbio*

We first compare the aggregate CMU/Facebook COVID-19 symptom survey responses to those from Burbio (Burbio, 2020). Burbio reports weekly measures for the percentages of public schools in each county in the country offering only virtual school, a combination of virtual and onsite school, or only onsite school (see Table 1). Burbio surveys a sample of 1,200 mostly large districts and then releases aggregate data at the county level, weighting by rough enrollment of the districts based on whether the districts are large, medium, or small (Burbio, personal communication; Harris et al., 2021). Data for the smallest counties are estimated by imputing from nearby counties (Harris et al., 2021). As was done to generate state aggregates from county percentages for the CMU/Facebook data, we aggregated the Burbio county-wide percentages to the state level by county population. All counties were used to calculate state-level percentages; however, only the largest 460 counties were used for county-level comparisons, because these counties had the most direct estimates (Burbio's methodology, 2020).

*MCH*

A second aggregate level data set comes from MCH (MCH, 2020). The MCH dataset contains district-wide designations of teaching method, which could be on premises, hybrid, online, unknown, other, or pending (MCH, 2020). The data is collected by reaching out to each district across the country (Harris et al, 2021). District teaching methods are not all updated at the same time, so MCH data was used from the January 31, 2021 update (see Table 1). Districts marked as private (n=60), Catholic (n=16), or missing (n=11) were removed for these comparisons, leaving a total of 15,541 public school districts in the dataset. County percentages of each schooling method were found by weighting each district by enrollment, and the same weighting by enrollment was done to aggregate up to the state level. All districts were used for state-level comparisons, and again only the largest 460 counties were included for county-level comparisons.

**Table 1.** Datasets used in comparison

| Source | Unit of Measure | Update Frequency | Dates Covered | Data Collection | Question | Sample Size | Weighting |
|---|---|---|---|---|---|---|---|
| CMU/ Facebook COVID-19 symptom survey | Household | Daily | 11/28/20 - 1/29/21 (excluding 12/22/20 -1/2/21) | Facebook platform-based survey | Do any of the following apply to any children in your household (pre-K–grade 12)? Going to in-person classes full-time; going to in-person classes part-time | 510,424 households (state-level) 354,458 households (county-level) | Facebook weights |
| Burbio | County | Weekly | 11/28/20 - 1/29/21 (excluding 12/18/20 -1/1/21) | 1,200 districts aggregated to county level, imputes for smaller counties | Percent of county offering: only virtual, combination, or traditional in-person | 3,142 counties (state-level) 460 counties (county-level) | To state: weighted by county population |

| MCH | District | Varies | 1/31/21 | Reached out to individual districts to get information | Teaching Method: on premises, hybrid, online, unknown, other, or pending | 15,537 districts  460 counties (county-level) | To state and to county: weighted by district enrollment |

*Other Sources*

There are a handful of other datasets that document school policies and household schooling behaviors during the pandemic, including the US Census Bureau PULSE survey (US Census Bureau, 2020), EdWeek (DATA, 2020), the UAS survey (USC, 2020), CRPE (CRPE, 2020), and JHU eSchool+ (Johns Hopkins Berman Institute, 2020). These sources were not included for comparison because some only included a select number of districts across the country (EdWeek and CRPE), questions were not easily comparable with the question of interest for this study (PULSE and JHU), or schooling-related questions were not asked during the November-December time period (UAS).

*County Level Factors*

Associations were examined between various county factors and differences in the percentage of households participating in onsite schooling between the data sources. County factors assessed included: unemployment rates in September 2020 (US BLS, 2020), population density from 2010 (Ykzeng, 2020), parental educational attainment from 2014-2018 (USDA ERS, 2020), 2016 mortality rate (Kochanek et al., 2017), 2010 rural versus urban classification (US Census Bureau, 2020), 2019 age breakdown of county (US Census Bureau, 2020), COVID-19 average cases and deaths over the entire period of November 28, 2020 to January 29, 2021

(length of comparison period for this study) (USA Facts, 2020), and other 2018 census data on race breakdown, population, health insurance, GINI coefficient, essential employment, computer and internet access, and more (Glenn, 2019). A full list of the variables included in the model can be found in the Results section to follow.

*Analyses*

Preliminary analyses involved exploratory graphs and correlations to assess the consistency between the three datasets at the county and state levels. Consistency was assessed primarily for the percentage of students attending any onsite school, which included fully onsite and hybrid (partially onsite). Secondary comparisons were done by breaking down the percentage into separate categories for fully onsite and hybrid; results for these are shown in the Appendix. These variables were compared on the state and county level. Our primary interest is in comparison of the results across the full time window; secondary analyses assessed week by week trends and found similar, but more noisy, relationships (see the Appendix). County-level pairwise differences were also calculated between the sources for the percentage of students with any onsite school.

Finally, county-level factors were merged with the school policy data to investigate the relationships between county-level factors and the difference in measures across sources for the most populous 460 counties in the US. Exploratory graphs were made to visualize the relationship between these variables and the difference between the CMU/Facebook source and MCH, and the CMU/Facebook source and Burbio, in terms of the percentage of students in the county with any onsite instruction. These visuals were fitted with LOESS curves, and weighted correlations were calculated for each comparison with the CMU/Facebook sample size as the weight (Bailey et al., 2018). Based on these weighted correlations, some county factors were also selected to be analyzed at the quartile level. For these factors, the distributions of the difference in percentage of onsite instruction were compared by quartile of the given county factor to see if different levels of the factor were associated with different percentage differences between the household reporting of onsite instruction and the county-wide policies.

**Results**

*Statewide Analyses*

Across the country during this time window, 47.2% of students had any onsite instruction according to the CMU/Facebook data source, 51.1% according to Burbio, and 50.8% according to MCH. Figure 1 shows the reported percentages by state and data source, showing fairly strong consistency during the time period considered. The correlations between the sources are high (Table 2), with the CMU/Facebook percentages having a 0.83 (p < 0.001) correlation with MCH percentages, and 0.92 (p < 0.001) with Burbio percentages, indicating high levels of consistency between the estimated percentage of any onsite school across states. The MCH and Burbio percentages were correlated at 0.80 (p < 0.001). The data sources follow a similar trend for each state.

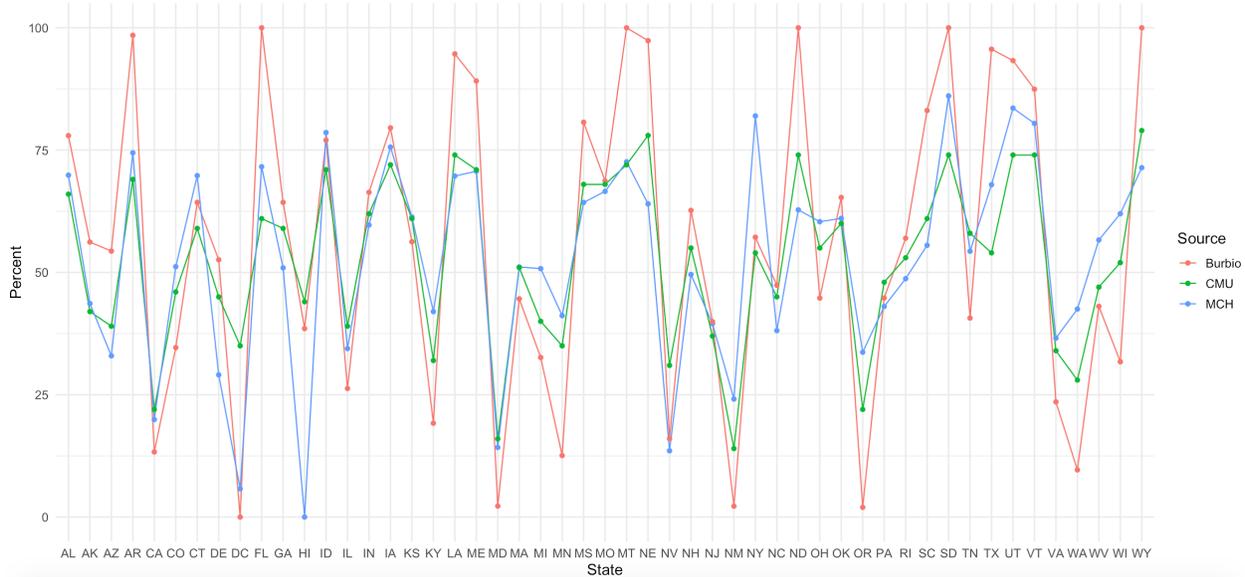

Figure 1. Percentage of students participating in any onsite instruction by state: November 28, 2020 - January 29, 2021

Table 2. Correlations between sources of measures of the state-level percentages of any onsite schooling

| Any Onsite School | MCH | Burbio | CMU/Facebook |
|---|---|---|---|
| MCH | 1 | 0.80 | 0.83 |
| Burbio | 0.80 | 1 | 0.92 |
| CMU/Facebook | 0.83 | 0.92 | 1 |

*County-level Analyses*

With over 3,000 counties in the United States and thus limited sample size in each county (ranging from 1 individual to 8,565 from the CMU/Facebook survey, with an average of 164 respondents per county), the 460 most populous counties were selected for county-level comparisons. In this sample the sample sizes from the CMU/Facebook survey ranged from 91 to 8,565 per county, with an average of 771. To visualize the consistency between CMU/Facebook and the other two sources, two scatterplots are presented (Figure 2). If full consistency is achieved, all points on the scatterplot would be on the y = x line. The scatterplot between Burbio and CMU/Facebook shows strong grouping around this line, while the MCH and CMU/Facebook data is slightly more spread. To confirm this, the percentage of students with any onsite school according to the CMU/Facebook survey was correlated at the county level 0.47 (p<.0001) with the MCH measure, and 0.87 (p<.001) with the Burbio measure (Table 3). Again, the MCH and Burbio data showed a lower correlation, 0.48 (p < 0.001).

Figure 2. Percentage of students with any onsite instruction by county for Burbio, MCH, and CMU/Facebook: November 28, 2020 - January 29, 2021

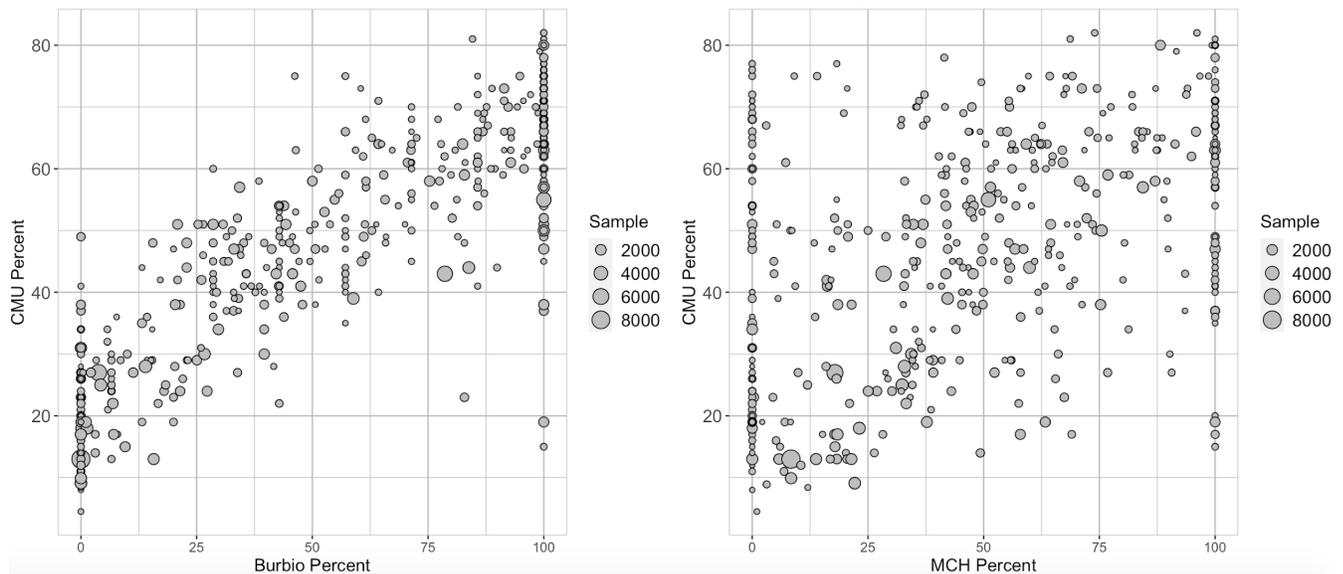

Table 3. Correlations between sources of county-level percentage of students with any onsite school

| Any Onsite School | MCH | Burbio | CMU/Facebook |
|---|---|---|---|
| **MCH** | 1 | 0.48 | 0.47 |
| **Burbio** | 0.48 | 1 | 0.87 |
| **CMU/Facebook** | 0.47 | 0.87 | 1 |

*Consistency across sources for hybrid schooling policies*

Secondary analyses similar to those above were performed on the data broken down into the percentages of fully onsite instruction and hybrid instruction. For the CMU/Facebook data, hybrid instruction was classified as going to school part-time; for Burbio, the variable was referred to as a combination of traditional onsite and virtual instruction; and for MCH, the variable was called hybrid learning. The results for these comparisons were less consistent across sources, with lower correlations for both categories of all onsite and hybrid (ranging from 0.11-0.78). The plots and correlations for these analyses are presented in the Appendix.

*Differences in measures at the county level*

Figure 3 displays the distributions of the difference between the CMU/Facebook and Burbio and MCH measures at the county level. The average difference in the county-level percentages among the most populous 460 counties between CMU/Facebook and Burbio was -6.29, meaning that CMU/Facebook reported about a 6 percentage point lower rate of any onsite schooling as compared to Burbio. The average difference in the county-level percentages between CMU/Facebook and MCH was -1.34, implying that CMU/Facebook shows about an 1 percentage point lower rate of any onsite schooling as compared to MCH.

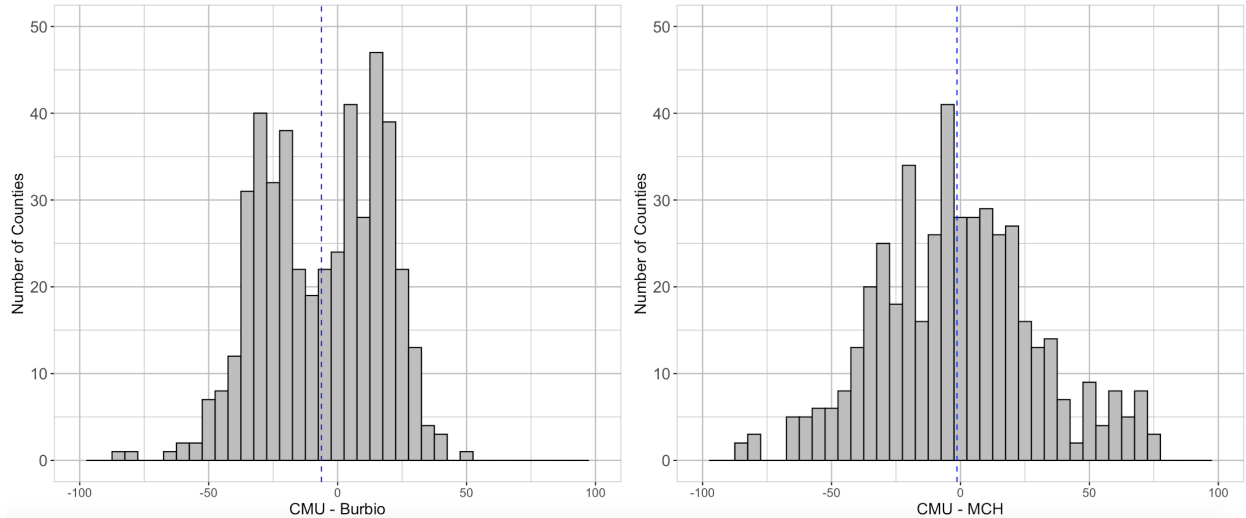

Figure 3. Differences between CMU/Facebook and Burbio and CMU/Facebook and MCH for measures of percentage of students with any onsite school by county

Next, scatterplots were created to investigate the relationships between differences between sources of onsite school percentages and county characteristics such as racial breakdown, unemployment rate, average COVID-19 cases, and economic factors. Weighted correlations were also calculated for these relationships, where weights were CMU/Facebook county sample size. Figures 4 and 5 display the scatterplots with weighted LOESS curves, and Table 4 shows the weighted correlations.

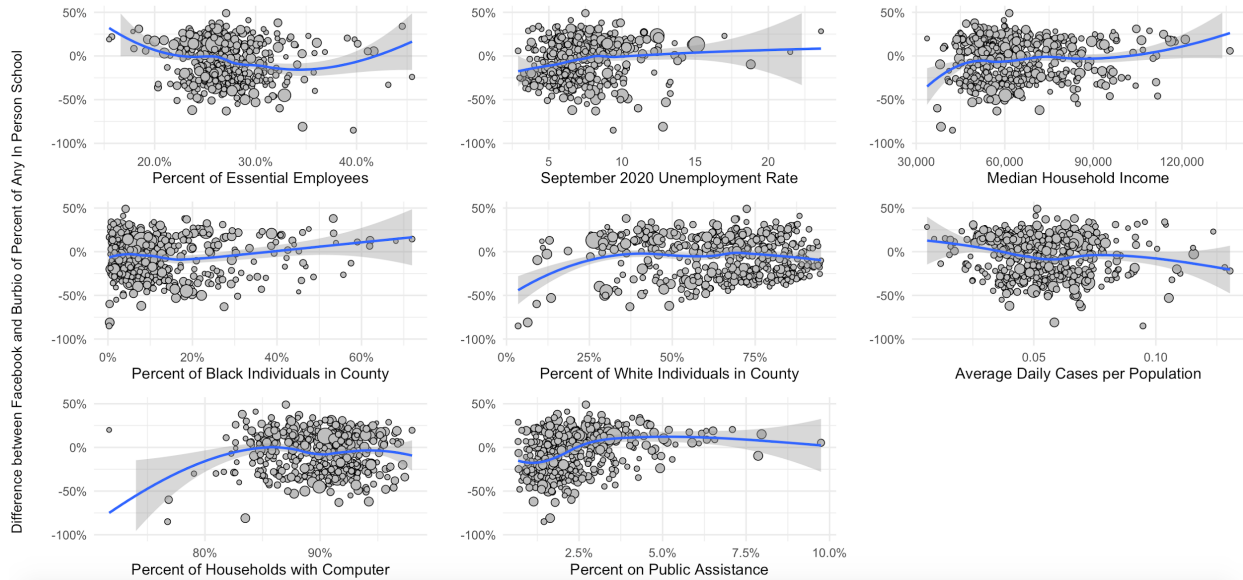

Figure 4. Relationship between county level characteristics and difference between CMU/Facebook and Burbio percentage of onsite school*

*Points represent a county where size is based on CMU/Facebook sample size, and LOESS curves are also weighted by sample size

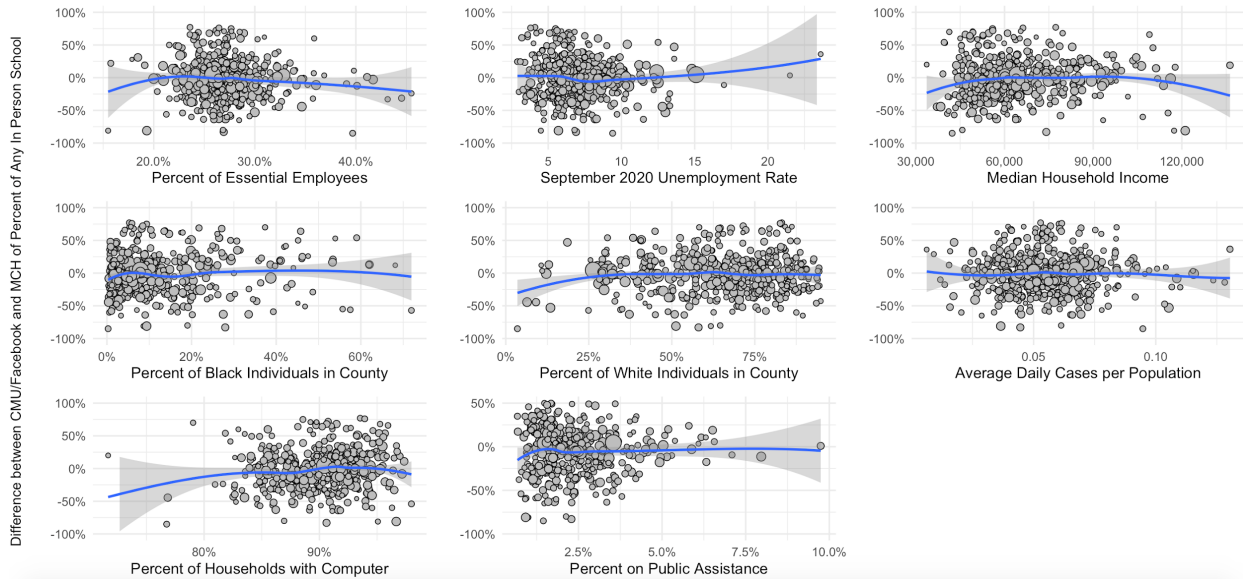

Figure 5. Relationship between county level characteristics and difference between CMU/Facebook and MCH percentage of onsite school*

*Points represent a county where size is based on CMU/Facebook sample size, and LOESS curves are also weighted by sample size

**Table 4. Weighted correlations between county level characteristics and difference between CMU/Facebook and Burbio and MCH percentage of onsite school**

| Variable | Weighted correlation with Difference between CMU/Facebook and Burbio percentage of onsite school | Weighted correlation with Difference between CMU/Facebook and MCH percentage of onsite school |
|---|---|---|
| Percent Receiving Public Assistance | 0.34 | -0.05 |
| Percent Essential Employees | -0.16 | -0.09 |
| September 2020 Unemployment Rate | 0.16 | -0.03 |
| Median Household Income | 0.11 | 0.06 |
| Average Cases per Population | -0.12 | -0.01 |
| Percent Black | 0.05 | 0.07 |
| Percent White | 0.05 | 0.04 |
| Percent of Households with Computer | 0.00 | 0.10 |

The weighted correlations in Table 4 show overall higher correlations between county factors and the CMU/Facebook versus Burbio difference compared to the CMU/Facebook versus MCH difference. Factors that are somewhat correlated with the difference in onsite school percentage between CMU/Facebook and Burbio include percentage receiving public assistance (e.g. SNAP, TANF) (r = 0.34), percentage of essential employees (r = -0.16), unemployment rate (r = 0.16), and average cases per population (r = -0.12) but overall the correlations are relatively small. Appendix Figure A.2 illustrates the difference between sources by quartiles of the variables showing the highest correlations. These generally show few differences, although a one-way ANOVA found a significant difference between the mean difference percentages by quartile of percentage receiving public assistance (p < 0.001), ranging from -15.7% in the lowest quartile to 4.9% for the highest quartile. A significant difference between the mean difference percentages by quartile was also found for unemployment rate (p = 0.002), ranging from -12.4% in the lowest quartile to -1.2% for the highest quartile. The quartile of percentage of essential employees also had a significant difference between the mean difference percentages. No other county-level variables showed significant differences across quartiles.

**Discussion**

There is substantial concern about the effects of onsite schooling on COVID-19 transmission, and more broadly about the effects of different school operational models on academic, social, and emotional outcomes of children, their families, and their communities. In order to generate evidence regarding these important questions, we need accurate information about the extent of onsite schooling that has been occurring throughout the pandemic and how it varies across households around the country. This paper has provided the first documentation of the consistency across different measures of onsite schooling across the country, comparing a large scale Facebook platform-based survey to two sources that collect data at the district and county levels.

The CMU/Facebook COVID-19 Symptom Survey is unique in its national scope and household-level data. This can allow examination of the experiences of individual households across the United States. It is reassuring that we find generally good consistency across sources in measures of the percentage of children receiving any onsite school at the state level and at the county level for large counties. More efforts should be put into collecting accurate data on schooling behaviors during the pandemic, especially at the household level and for smaller, rural counties. However, these results give some reassurance that at broad levels we can use existing household-level data to understand the extent of onsite schooling during the pandemic.

Notably, it is harder to distinguish between the extent of onsite schooling in terms of fully onsite compared to part-time onsite or a hybrid model. Part of this difficulty might be due to the various possible definitions of "hybrid" schooling, which can include districts where each individual school operated on a partially online/partially onsite format, or districts where some schools (like elementary schools) were fully onsite or hybrid but other schools (like high schools) were fully online (Harris et al., 2021). Therefore, each district, each country, and/or each source (MCH, Burbio, CMU/Facebook) could have a different definition of what hybrid means. Additionally, the difficulty of distinguishing could also be due to school policies changing between hybrid and all onsite over time. This signals that it may be difficult to understand the nuances of the extent of onsite schooling, at least using these data sources.

These findings support the consistency between the datasets, and they also support the possible hypothesis that at the household level, students are being sent to school slightly less often than the official school or district policies would imply - although districts might be open for onsite school, not all families will elect to send their children. This is seen in this data with the overall lower rates of the percentage of students with onsite schooling reported in the CMU/Facebook data as compared to Burbio and MCH.  Furthermore, as shown in Figure 2 comparing the county level percentages of onsite instruction, there are many counties at 100% onsite according to Burbio or MCH but less than 100% according to CMU/Facebook. This indicates that for many counties across the country, even if the policy of the county is all onsite, many families "opt out" and do not send their children to school onsite. Although private schools could be playing some small role in this distinction since CMU/Facebook includes private schools but the other two sources do not, this result is still striking and emphasizes why the individual household level data provided by CMU/Facebook is important to show the decisions that families are making.

In examining whether the difference across sources varies across counties with different characteristics, few relationships were found. There were some characteristics, however, that had a moderate correlation with the difference across the CMU/Facebook survey and the Burbio dataset, in particular the percent receiving public assistance, the unemployment rate, and the percentage of essential employees in the county. As demonstrated by the analyses at the quartile level, lower quartiles of unemployment rate had more negative average percentage difference in onsite reporting, meaning that counties with lower unemployment rates had larger differences in reporting across sources, and in particular more families in the CMU/Facebook data reporting no onsite school, i.e., a higher "opt out" rate. A similar relationship occurred for quartiles of percentage receiving public assistance, in that lower percentages of the county receiving public assistance were associated with more families in the CMU/Facebook data reporting no onsite school. These county-level factors can be further explored to determine more about how they relate to individual families' decisions to send their children to school or not.

While this study is the first that assesses consistency between data sources and helps motivate future research on the effects of school policies during the COVID-19 pandemic, a limitation of the study is that it gives only a snapshot of the levels of consistency, examining late Fall and early Winter 2020/2021. Other patterns may have been seen in early Fall 2020 and later into 2021.  An additional limitation is that the CMU/Facebook survey does not distinguish the type of schooling (e.g., private, private, charter), whereas Burbio and MCH both focus on public schools and school districts. The CMU/Facebook survey also includes Pre-Kindergarten children, while Burbio and MCH do not necessarily include all Pre-K schools.  Due to sample size limitations the county-level analyses focused on the largest counties in the country; implications for small counties are unclear.

With so many unanswered questions about and a lack of understanding of the consistency of K-12 schooling policies across the country during the pandemic, this paper serves to highlight the data that exists and the consistency between sources. The CMU/Facebook COVID-19 Symptom Survey's information on household experiences across the entire country will allow for many important questions to be answered on a detailed level, especially with the deeper understanding gained by this paper of the consistency of this survey with other nationwide data sources on educational policies and experiences collected at the district and county-level.

**Acknowledgements**

This research is partially based on survey results from Carnegie Mellon University's Delphi Group.  We also thank Burbio and MCH for data access.

**Appendix**

*A.1 Hybrid Schooling Comparisons*

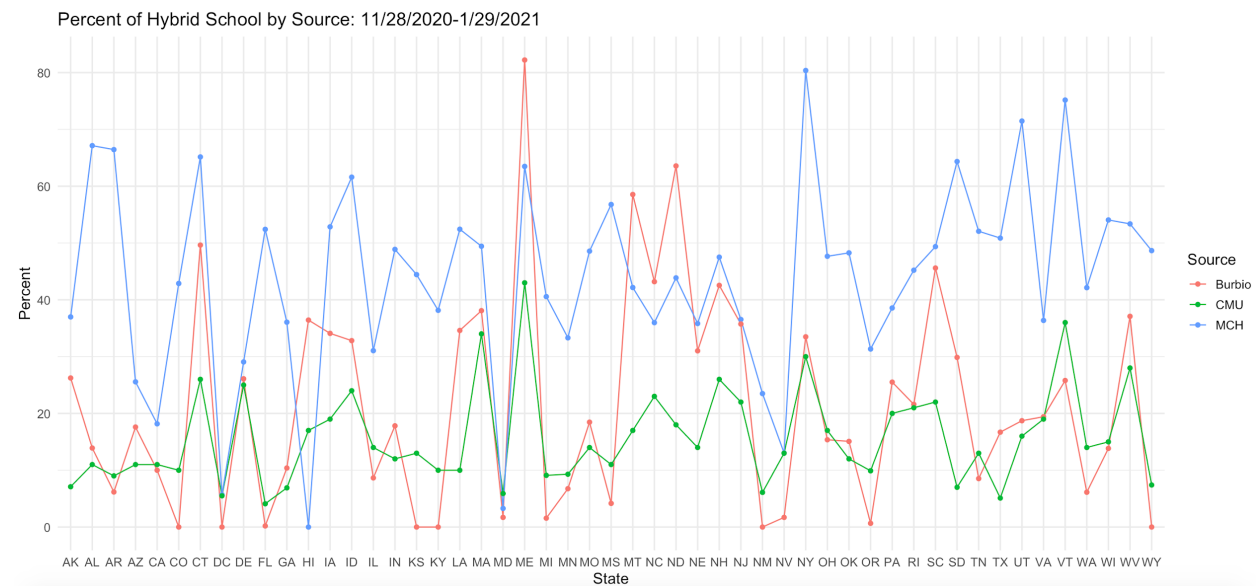

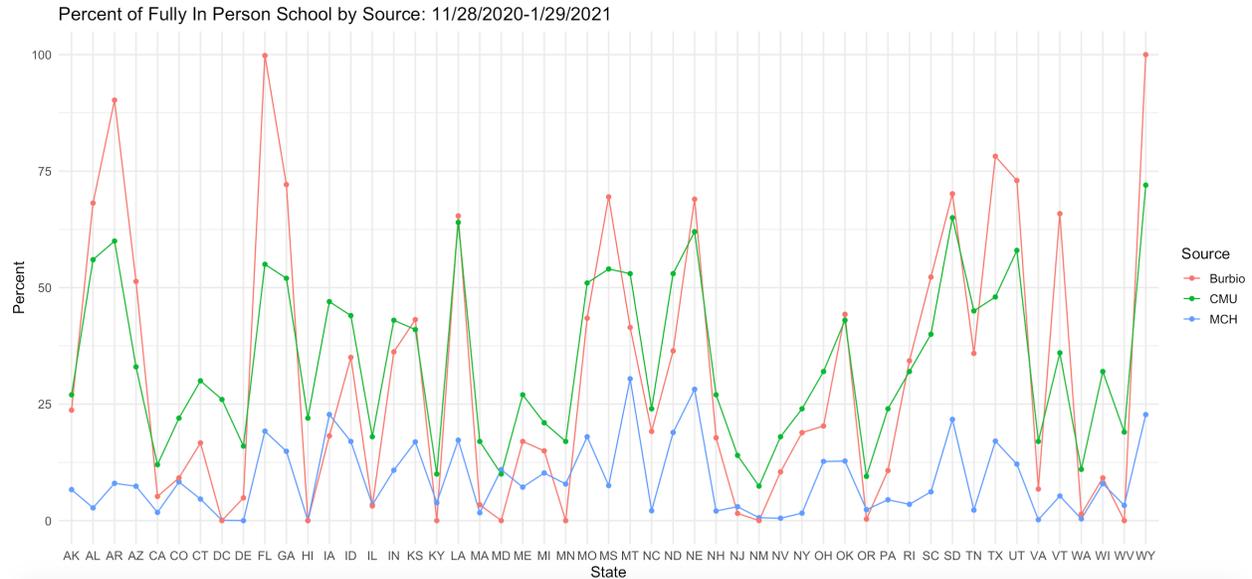

Table A1. Correlations between sources of percent of hybrid school by state*

| Any Onsite School | MCH | Burbio | CMU |
|---|---|---|---|
| MCH | 1 | 0.29 | 0.38 |
| Burbio | 0.29 | 1 | 0.72 |
| CMU/Facebook | 0.38 | 0.72 | 1 |

*Hybrid had various meanings, for MCH it was called hybrid; for Burbio, combination of virtual and onsite; for CMU/Facebook, part-time

Table A2. Correlations between sources of percent of hybrid school by county

| Any Onsite School | MCH | Burbio | CMU |
|---|---|---|---|
| MCH | 1 | 0.11 | 0.15 |
| Burbio | 0.11 | 1 | 0.78 |
| CMU/Facebook | 0.15 | 0.78 | 1 |

*A.2 County-Level Factors by Quartile*
*an ANOVA indicated significant difference between means across quartiles

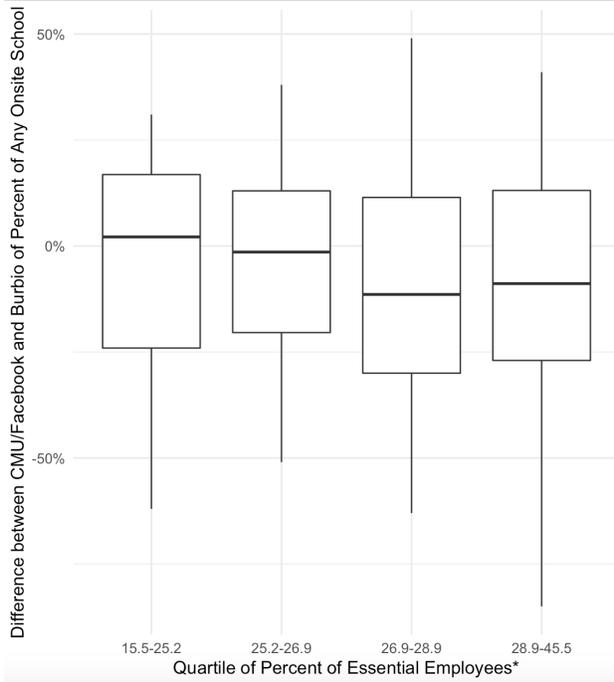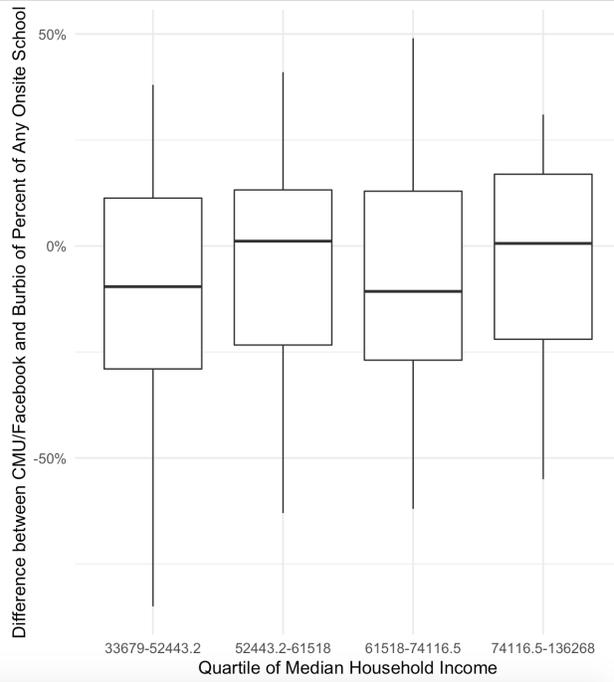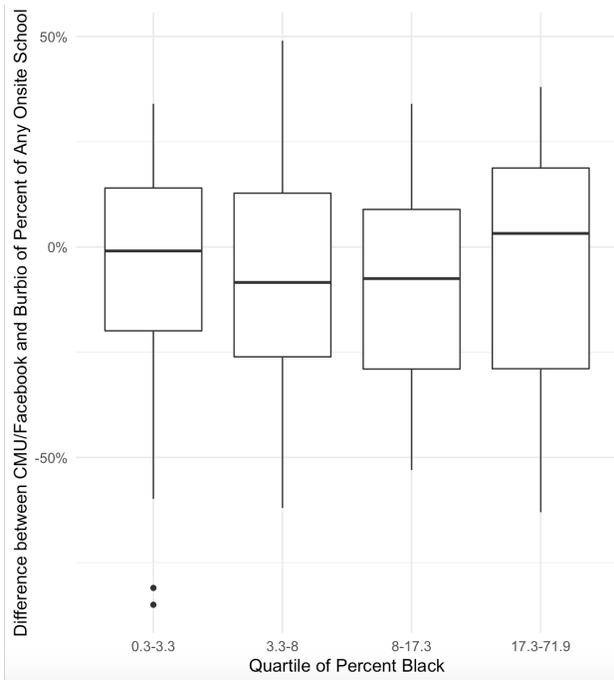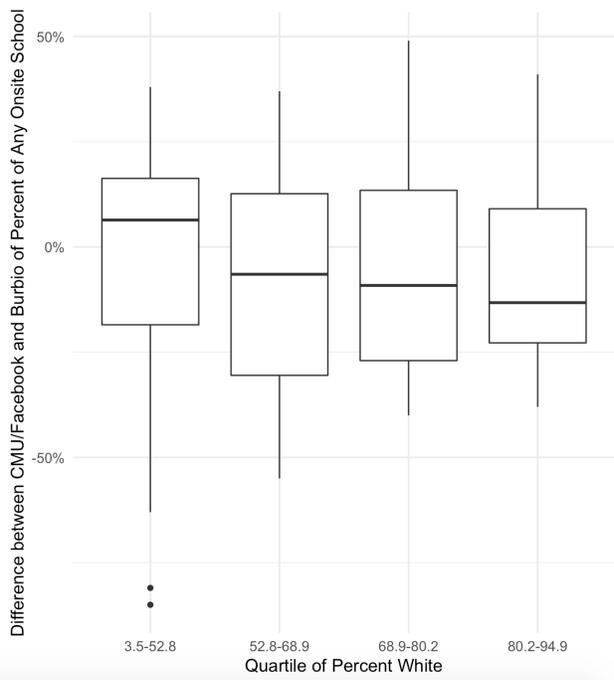

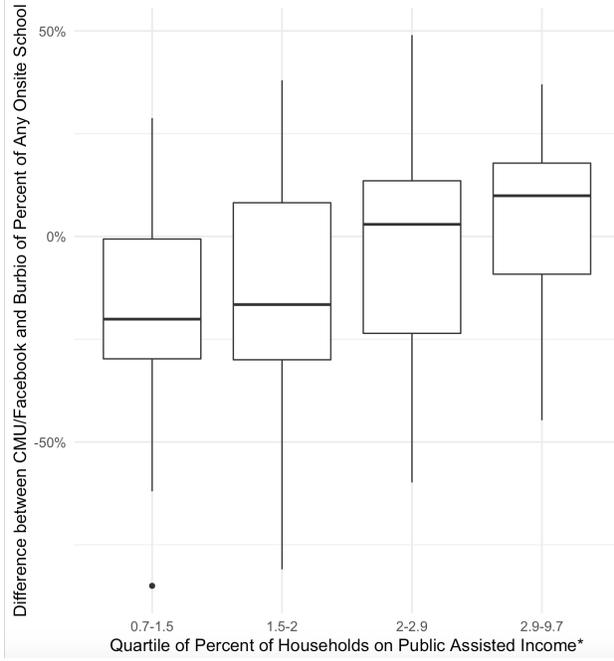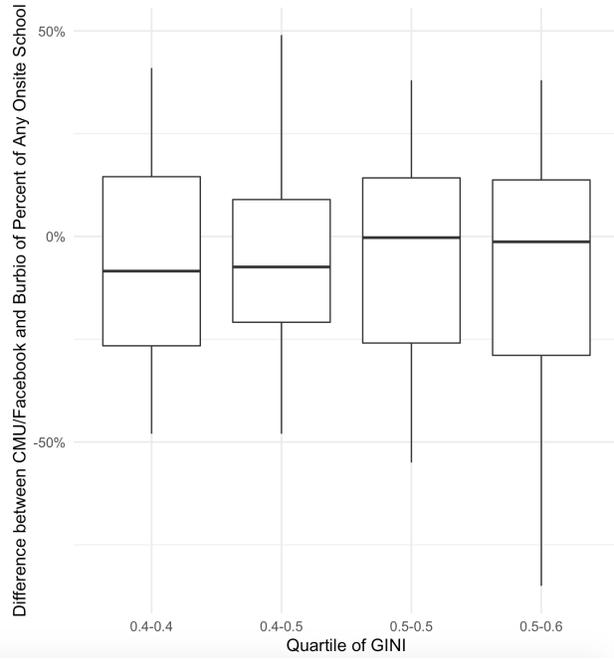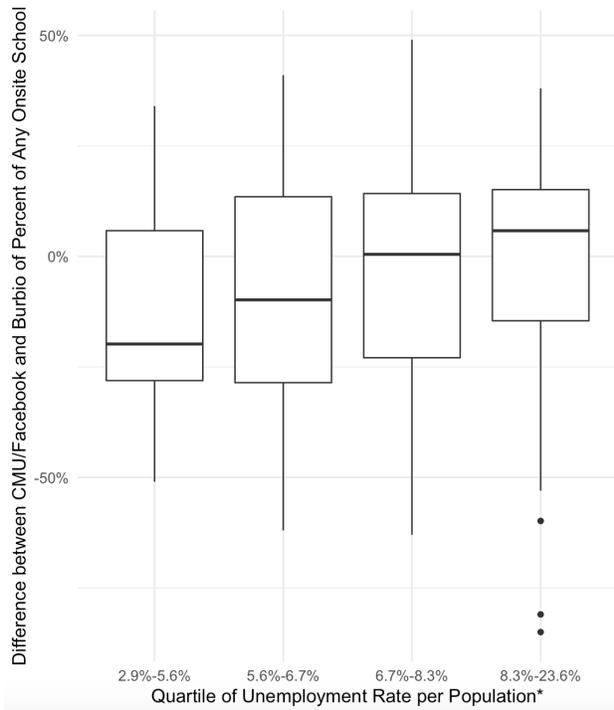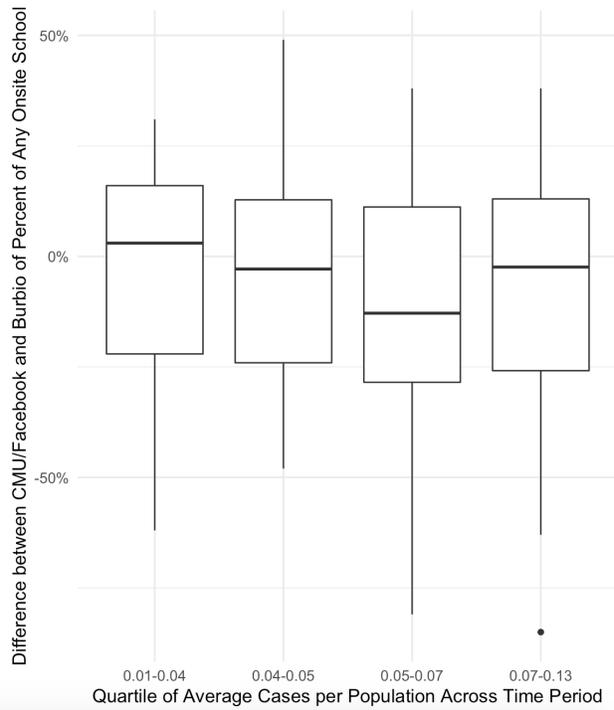